**Title**: Memristor variability and stochastic physical properties modeling from a multivariate time series approach

**Authors:** F. J. Alonso[1], D. Maldonado[2], A. M. Aguilera[1], J.B. Roldán[2]

**Address:**

[1]Departamento de Estadística e Investigación Operativa. Universidad de Granada. Facultad de Ciencias. Avd. Fuentenueva s/n, 18071 GRANADA, Spain.

[2]Departamento de Electrónica y Tecnología de Computadores. Universidad de Granada. Facultad de Ciencias. Avd. Fuentenueva s/n, 18071 GRANADA, Spain. Email: jroldan@ugr.es

**Abstract:**

A powerful time series analysis modeling technique is presented to describe cycle-to-cycle variability in memristors. These devices show variability linked to the inherent stochasticity of device operation and it needs to be accurately modeled to build compact models for circuit simulation and design purposes. A new multivariate approach is proposed for the reset and set voltages that accurately describes the statistical data structure of a resistive switching series. Experimental data were measured from advanced hafnium oxide based devices. The models reproduce the experiments correctly and a comparison of the multivariate and univariate approaches is shown for comparison.

# I.-INTRODUCTION

Memristors in general, and in particular those linked to resistive switching (RS) operation, are being scrutinized both by the Academy and the industry to explore their outstanding potential in the non-volatile memory realm [1, 2]. These nanometric electron devices show inherent stochasticity in their operation that leads to different flavors of variability: device-to-device, connected to technological differences that might show up in the fabrication process; and cycle-to-cycle, associated to the random physical mechanisms behind their operation. This latter phenomenon can hinder the memristor massive industrial exploitation for non-volatile memory circuits; however, on the contrary, it could pave the way to employ these devices as entropy sources in the context of hardware cryptographic applications. In this respect, recent interesting works have been published [3-5].

Memristors are also key actors in the neuromorphic circuit playground since they can easily mimic biological synapses that facilitate hardware neural networks fabrication [6-10]. This research topic is drawing attention at all levels due to the need for efficient, low power and fast artificial intelligence solutions to process the huge data amounts generated currently by social networks, internet-of-things devices, e-commerce… Memristor fabrication is fully compatible with Complementary Metal-Oxide-Semiconductor (CMOS) technology and, consequently, it can be used to push forward chips devoted to neuromorphic computing, a term coined by Carver Mead [11] when the basis for hardware neural network implementation was laid out in the late eighties.

There are many types of devices, since their structure (two electrodes with a dielectric in between) can be fabricated with a wide range of materials, including 2D materials, such as graphene oxide [11], h-BN [1], etc. Although RS phenomena for some devices are relatively well-known, there is a long way to go to fully unveil the physics behind resistive switching and the corresponding variability; on this issue, many experimental, modeling and simulation analysis have been published in the last few years [2, 13-25]. Electronic design automation (EDA) tools are essential for circuit design in electronics; therefore, the maturity of a technology is linked to the availability of physical simulation tools and compact models. From the modeling view point, different approaches can be found in the literature for memristors [20, 18, 14, 21, 26, 27]. Variability issues need to be addressed in the modeling context by the scientific community since, at this moment, they are key for circuit design. It is interesting to point out that neuromorphic applications are more tolerant to the inherent stochastic nature of resistive switching (RS) [28, 29]. In fact, this stochasticity can be used to build true random number generators in the context of integrated circuit cryptography [3].

Most memristors are based on resistive switching of filamentary nature [1, 2]. In these devices, the creation (set process) and destruction (reset processes) of conductive filaments (CFs) allows RS. Both, reset and set events are linked to random processes connected to physical mechanisms: diffusion, redox and nucleation…, both for valence change memories and for conductive bridge RAMs [25, 17, 2, 23, 21, 1, 19, 20]. That is why RS stochasticity, experimentally clear in terms of device resistance and set and reset voltages, produce the cycle-to-cycle (CTC) variability in long RS series (the series consists of repetitive cycles of set and reset events).

Among the different strategies employed to deal with the statistical analysis of CTC variability, the use of Weibull distribution (a weakest-link type distribution employed in reliability studies [30]) can be counted [31, 32, 2]. One of the drawbacks of Weibull distribution to analyse memristors RS data is the assumption of independent times to failure (voltages to reset or to set processes in our case). As reported in Roldán et al. [33], stochasticity linked to successive observations could show data dependence (CFs are reformed using the remnants of previous ones as starting point). Taking into consideration this fact, we stepped forward and changed the modeling approach into the time series playground [33, 34, 35, 36]. Time series analysis (TSA) has been successfully employed to characterize RS in different device types: conductive bridge RAMs with Ni electrodes and $HfO_2$ as a dielectric [33] and graphene oxide based devices [12]. Indeed, memristor CTC cycle-to-cycle variability can be modeled by TSA, assuming that the experimental measurements (consecutive reset and set voltages) were obtained successively in time for a RS long series.

G.U. Yule introduced modern TSA to model the temporal movement of a pendulum [37]. This modeling technique allows the description of RS inertia in memristors. In our first approach, univariate TSA was employed for reset and, separately, for set processes. However, in this manuscript we tackle with a multivariate strategy for the first time that is more appropriate since the set process starts making use of the remnants of the CFs destroyed in the previous cycle (a reset process). In this respect, the reset voltage has dependencies on the values of neighbouring cycles (previous cycles reset voltages) and also on previous cycles set voltages, keeping the system

"memory" in a long RS series. As explained in [33], this means that consecutive RS cycles can be correlated in a series. Therefore, the concept of autocorrelation has to be accounted for. The time series stationarity was also studied in the context of memristors in [33] as a needed tool for TSA [34, 35]. In this new TSA modeling approach, we deal with a higher numerical intricacy in order to obtain a more truthful description; in this sense, and comparing with the monovariate TSA presented in [33], we face a different scenario in the usual trade-off of accuracy and complexity within the modeling paradigm.

Prior to the application of a multivariate strategy, the causality between the univariate series has to be studied [37]. This property checks for the improvement obtained when the modeling of one variable (the set or reset voltage, as we shall show below in Figure 1) includes, in addition to past values of the same variable (univariate model, as reported in [33]), values of other variables; i. e. a reset voltage model would include past values of the reset and set voltages. In the case of a multivariate approach, the model is selected using the extended correlation matrix. This matrix provides the p-values of multivariate Ljung-Box statistics of a vector series [39]. The model proposed here will be used for forecasting more accurately the values of set and reset jointly.

The devices studied here are obtained from bipolar valence change memories based on a TiN/Ti/HfO$_2$/W/Ti stack. The Ti layer in the top electrode works as an oxygen ion reservoir [1, 20]. The manuscript is organized as follows: in section II the fabricated devices and measurement process are described, the statistical procedure is given in section III and the main results and discussion are included in section IV. Finally, in section V we draw the main conclusions up.

## II.-FABRICATED DEVICES AND MEASUREMENT

The memristors top electrode is made of a bi-layer stack (200 nm TiN/10 nm Ti), the bottom metal consists of a 50 nm thick W layer over a Ti layer. The dielectric (a 10 nm $HfO_2$ layer) was deposited by ALD. More details on the fabrication process can be found in [40]. For the configuration of the measurement set-up, the bottom electrode was grounded and a voltage signal with a (0.8V/s) ramp was applied to the top electrode (the voltage step was 0.01V). A RS series of 1000 cycles (consecutive reset and set events) were carried out after a forming process, see Figure 1. The devices were square-shaped with an area of 15×15 µm$^2$.

The set ($V_{set}$) and reset ($V_{reset}$) voltages were obtained following the numerical procedures in line with [20, 16], see Figure 1.

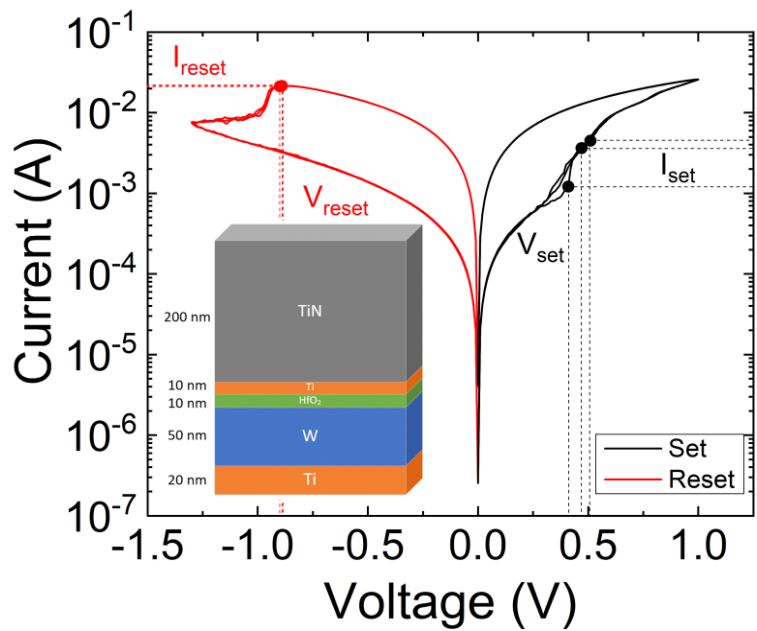

**Figure 1.** Experimental I-V curves for 3 set/reset consecutive cycles showing $V_{set}$, $V_{reset}$, $I_{set}$ and $I_{reset}$. The layer stack scheme for the devices under study is shown in the inset.

The set ($I_{set}$) and reset ($I_{reset}$) currents are the current values where $V_{set}$ and $V_{reset}$ take place. The latter parameters have been plotted in Figure 2 for the cycles considered in the measured RS series.

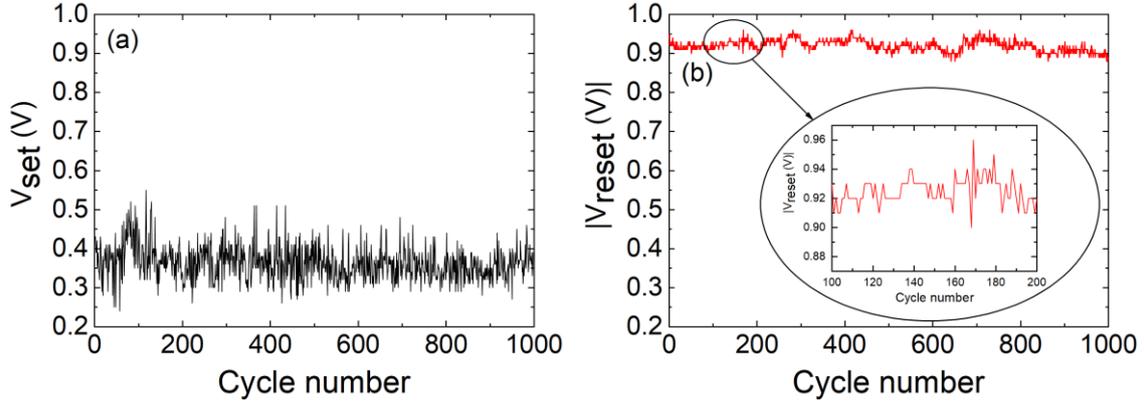

**Figure 2.** Experimental values of $V_{set}$ (a) and $V_{reset}$ (b) versus cycle number for a series of continuous RS cycles under ramped voltage stress for memristors based on TiN/Ti/HfO$_2$/W/Ti stacks.

## III.-STATISTICAL METHODOLOGY

## III.A.-MULTIVARIATE TSA DESCRIPTION

In the time series analysis we are presenting here, $V_{set}$ and $V_{reset}$ of the devices under study are modeled. Other RS parameters could also be modeled such as the reset and set currents (see Figure 1). As already stated in [33-35], finding the model order is key; i. e. how many set and reset voltage values from previous cycles we need to accurately forecast the current cycle (see Equation 1). The model order depends on the physics underlying RS; nevertheless, all the information for our modeling is extracted from just the measured data and their probabilistic structure, no other assumptions have to be considered in our analysis.

The TSA numerical procedure also helps to find the weights ($\Phi_1 \ldots \Phi_p$) of the autoregressive (AR) model we are seeking (see Equation 1). Henceforth, for simplification purposes, we work with the absolute value of $V_{reset}$.

$$V_{reset_t} = \Phi_0 + \Phi_1 V_{reset_{t-1}} + \Phi_2 V_{reset_{t-2}} + \cdots + \Phi_p V_{reset_{t-p}} + \varepsilon_t \qquad (1)$$

The term $\varepsilon_t$, accounts for the model error (the difference between the measured value and the modeled value), it is known as the residual.

Equation 1 shows an autoregressive model, but sometimes a more complicated expression is needed to predict experimental data; e. g. an autoregressive moving average (ARMA) model that includes AR and moving average (MA) parts [34-35]. MA models are a linear combination of past residuals [34-35]. The general expression ARMA(p,q) model is given by the following expression,

$$V_{reset_t} = \Phi_0 + \Phi_1 V_{reset_{t-1}} + \cdots + \Phi_p V_{reset_{t-p}} - \theta_1 \varepsilon_{t-1} + \cdots - \theta_q \varepsilon_{t-q} + \varepsilon_t \qquad (2)$$

The algebraic equations above can be used for device compact modeling [33]. The mathematical expressions of a compact model implemented in a circuit simulator allow to account for the device behaviour at the circuit level and, consequently, circuit designers can have an invaluable tool to carry out their tasks. So, the developments to be presented here should be contextualized within circuit simulation applications and the great added value behind this type of models should be noticed, since the inherent cycle-to-cycle variability of these devices can be accounted for.

Both, Equations 1 and 2 show a univariate approach, which was presented in depth in [33]. However, successive reset and set processes could be connected in a RS series, therefore the appropriateness of a multivariate model is clear and physically reasonable. We deal with this issue here; in particular, we deal with $V_{reset}$ and $V_{set}$ values in expressions similar to Equations (1) and (2) but accounting for crossed dependencies (i.e., we will use mathematical expressions for $V_{reset}$ and $V_{set}$ that include in both cases previous values of the two different magnitudes; in this manner, $V_{reset}$ depends on previous values of $V_{reset}$ and $V_{set}$ and the other way around).

**III.B.-CAUSALITY STUDY**

Causality studies shed light on the goodness of a multivariate model implementation. This means the evaluation of a certain variable (e. g. the reset voltage) regression series involving past values of another variable (e. g. the set voltage) to scrutinize if there is an improvement in the representation model; thus is, if the model depicted below in Equation 3,

$$V_{reset_t} = \Phi_0 + \Phi_1 V_{reset_{t-1}} + \cdots + \Phi_p V_{reset_{t-p}} + \alpha_1 V_{set_{t-1}} + \cdots + \alpha_p V_{set_{t-p}} + \varepsilon_t \quad (3)$$

presents a significant contribution compared to the model described by Equation 1. A direct analysis on the measured values seems straight forward; nevertheless, a better manner to evaluate this issue stands upon a comparison of the residuals of the corresponding univariate models. Thus, using a model ARIMA(0,1,2) for the set voltage and ARIMA(0,1,1) for the reset voltage, which are the more compact options we found for the univariate approaches, the Granger test [38] output indicates a non-negligible causality (statistical connection) among these variables. A p-value of 0.0152 is obtained for the decision between Equation 1 and 3 for $V_{reset}$, and in a parallel calculation a p-value of 0.00364 for the equations for $V_{set}$ (notice that a p-value lower than 0.05 indicates that a model based in Equation 3 is more appropriate than a model linked to Equation 1).

Another tool to check causality consist of the Cross Correlation Function (CCF) among the residuals of the univariate models. The cross correlation function between $V_{set}$ and $V_{reset}$ is given by the following expression,

$$CCF(k) = corr(V_{set_{t+k}}, V_{reset_t}) = \frac{Cov(V_{set_{t+k}}, V_{reset_t})}{\sqrt{Var(V_{set_{t+k}}) \, Var(V_{reset_t})}} \qquad (4)$$

where Cov stands for the covariance and Var for the variance [33]. A significant value of CCF(k) compared to the threshold bounds (calculated as $\pm\frac{1.96}{\sqrt{n}}$, where n is the number of measured data, i. e. the number of cycles in our case [Brockwell02]. We have employed 1000 cycles of consecutive set and reset I-V curves in the RS series, therefore n=1000 and the corresponding threshold bounds are ±0.06198) for k<0 implies that including $V_{set}$ values would model better $V_{reset}$ because the past values of the set voltage are correlated with the actual value of the reset voltage (a significant value of $corr(V_{set_{t-l}}, V_{reset_t})$ implies that the $V_{set}$ past improves $V_{reset}$ modeling). On the contrary, if a significant CCF(k) value is obtained for k>0, it implies that $V_{set}$ can be better explained relating the variable also with $V_{reset}$; (a significant value of $corr(V_{set_{t+l}}, V_{reset_t}) = corr(V_{set_t}, V_{reset_{t-l}})$ implies that the $V_{reset}$ past values improve $V_{set}$ modeling). As noted previously, the cross correlation function (CCF, as defined in (4)) points out better results when it is computed over the univariate residuals, removing the intrinsic correlation over each variable.

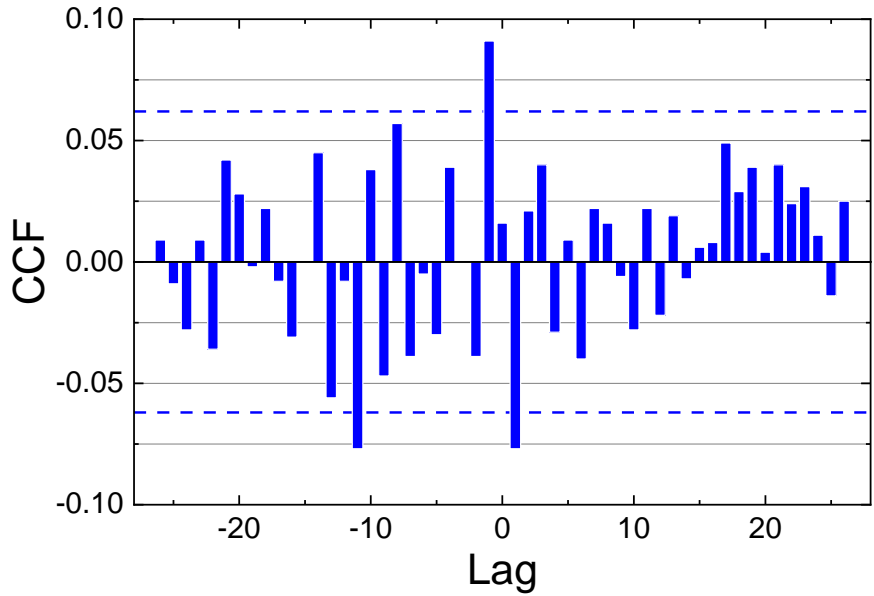

**Figure 3.** Cross correlation function (CCF) for the $V_{set}$ and $V_{reset}$ univariate time series residuals. The CCF minimum threshold bounds for the experimental data we are employing are ±0.06198, shown with dashed lines.

Only the values corresponding to the lags k=-11, k=-1 and k=1 pass the threshold bound (0.06198) in Figure 3. This result implies that both variables $V_{set}$ and $V_{reset}$ show a causality relationship, and consequently a multivariate model will perform better than univariate models.

### III.C.-MODEL ORDER SELECTION

The ARMA multivariate model (or vector model, VARMA) is described by the following equation,

$$\begin{pmatrix} V_{set} \\ V_{reset} \end{pmatrix}_t = \begin{pmatrix} \Phi_{set} \\ \Phi_{reset} \end{pmatrix} + \sum_{k=1}^{p} A_k \begin{pmatrix} V_{set} \\ V_{reset} \end{pmatrix}_{t-k} + \sum_{k=1}^{q} B_k \begin{pmatrix} \varepsilon_1 \\ \varepsilon_2 \end{pmatrix}_{t-k} + \begin{pmatrix} \varepsilon_1 \\ \varepsilon_2 \end{pmatrix}_t \qquad (5)$$

where $\Phi_{set}$ and $\Phi_{reset}$ are constant values, and $A_k$ and $B_k$ are 2x2 matrices. The section objective is to develop a procedure to decide if the constants $\Phi$ need to be included in the model, and how many matrices A and B have to be estimated.

There are many criteria to select the model [41]. The most frequently used are based on the likelihood and the needed number of parameters over some selected model candidates. These criteria tend to overestimate the model. For example, in our case, for VAR models (in this case B matrices are not employed) the value of p is equal to 10 using the AIC criterion (Akaike's Information Criterion) [41].

Another criterion is based on the use of extended cross-correlation matrices. This criterion computes the p-values of the multivariate Ljung-Box statistics [39]. The statistics computed are a multivariate version of Ljung-Box test that evaluates if the first "L" values of the ACF function are equal to 0 or not. In this context, to determine the parsimonious (the simpler and with the lower number of terms) model order, it must be used a p-values table of Extended Cross-correlation Matrices. This table can be calculated using the R-function ECCM [39]. In particular, in the p-values table, it must be found a triangle corner with all the elements higher than 0.05 (see Table I). Table I suggests the possibility of models VARMA(0,2) (yellow) or VARMA(1,1) (green) using the differenced time series, i. e. a time series accounting for values, once a variable change has been taken into consideration (for example, $DV_{set_t} =$ $V_{set_t} - V_{set_{t-1}}$). This variable change is performed when the original data series is not stationary. Many times a simple variable change like this leads us to a stationary data series, a needed condition to be able to develop a time series model [33].

```
Column: MA order                          Column: MA order
Row   : AR order                          Row   : AR order
       0      1      2      3      4      5      6              0      1      2      3      4      5      6
0  0.0000 0.0072 0.2323 0.2057 0.3118 0.2656 0.3079       0  0.0000 0.0072 0.2323 0.2057 0.3118 0.2656 0.3079
1  0.0000 0.2383 0.5574 0.8835 0.9349 0.5406 0.7682       1  0.0000 0.2383 0.5574 0.8835 0.9349 0.5406 0.7682
2  0.0000 0.3810 0.9405 0.9589 0.3769 0.7869 0.6352       2  0.0000 0.3810 0.9405 0.9589 0.3769 0.7869 0.6352
3  0.0000 0.9057 0.9991 0.7705 0.9207 0.9348 0.9955       3  0.0000 0.9057 0.9991 0.7705 0.9207 0.9348 0.9955
4  0.0000 0.9810 0.9989 0.9989 0.9864 0.8998 0.9997       4  0.0000 0.9810 0.9989 0.9989 0.9864 0.8998 0.9997
5  0.0010 0.9991 1.0000 1.0000 1.0000 1.0000 0.9131       5  0.0010 0.9991 1.0000 1.0000 1.0000 1.0000 0.9131
```

**Table I.** p-values table of Extended Cross-correlation Matrices. Triangles with all elements greater than 0.05 suggest the type of model to be used (yellow corresponds to a VARMA(0, 2), green corresponds to a VARMA(1,1)).

The likelihood obtained for the modeling approach described above is very similar, as can be deduced by the residuals correlation matrices. The selected model in the first case is of VARMA(1,1) kind, providing the estimated model expressed in Equation 6,

$$\begin{pmatrix} DVset \\ DVreset \end{pmatrix}_t = \begin{pmatrix} 0.102983 & -0.550028 \\ 0.019652 & -0.085120 \end{pmatrix} \begin{pmatrix} DVset \\ DVreset \end{pmatrix}_{t-1} + \begin{pmatrix} \varepsilon_1 \\ \varepsilon_2 \end{pmatrix}_t - \begin{pmatrix} 0.921835 & 0.201058 \\ 0 & -0.662339 \end{pmatrix} \begin{pmatrix} \varepsilon_1 \\ \varepsilon_2 \end{pmatrix}_{t-1} \quad (6)$$

This vector equation is equal to the following single equations:

$$\begin{aligned} Vset_t &= 1.102983\ Vset_{t-1} - 0.102983\ Vset_{t-2} \\ &\quad - 0.550028\ (Vreset_{t-1} - Vreset_{t-2}) - 0.921835\ \varepsilon_{1_{t-1}} \\ &\quad - 0.201058\ \varepsilon_{2_{t-1}} + \varepsilon_{1_t} \\ Vreset_t &= 0.91488\ Vreset_{t-1} + 0.08512\ Vreset_{t-2} \\ &\quad + 0.019652\ (Vset_{t-1} - Vset_{t-2}) + 0.662339\ \varepsilon_{2_{t-1}} + \varepsilon_{2_t} \end{aligned} \quad (7)$$

The variance of the errors is given by 0.001739279 for $\varepsilon_1$ and 0.00008128950 for $\varepsilon_2$.

The corresponding univariate models are given by the expressions below (Equations 8)

$$Vset_t = Vset_{t-1} - 0.8136\, \varepsilon_{1_{t-1}} - 0.0965\, \varepsilon_{1_{t-2}} + \varepsilon_{1_t}$$

$$Vreset_t = Vreset_{t-1} - 0.7095\, \varepsilon_{2_{t-1}} + \varepsilon_{2_t}$$

(8)

The error variance is higher than in the multivariate case. Therefore, the multivariate modeling provides a better fit to the measured values.

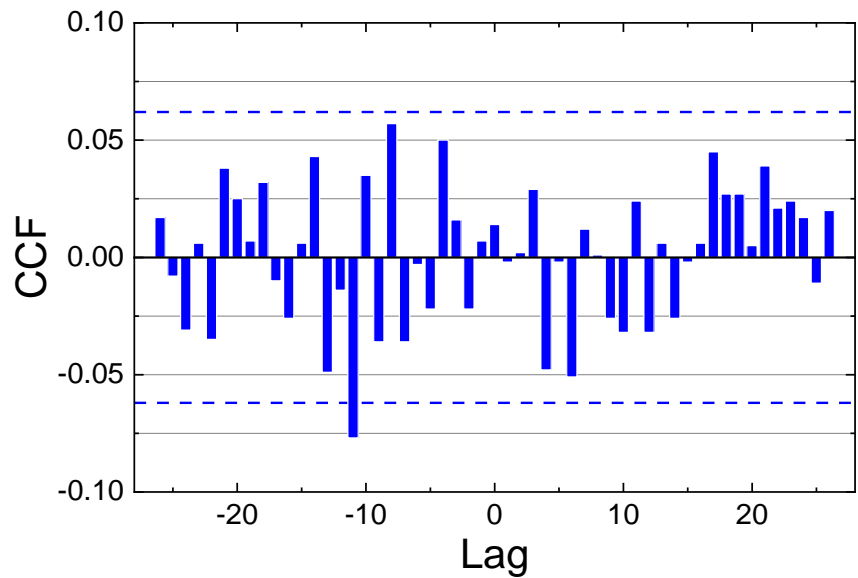

**Figure 4.** Cross correlation function for residuals of the VARMA(1,1) model for $V_{set}$ and $V_{reset}$. The CCF minimum threshold bounds for the experimental data we are employing are ±0.06198, shown with dashed lines.

The ACF and PACF for the residuals show that they are uncorrelated. All the values of the cross-correlation function (Figure 4) are between the threshold bounds (only the value -11 cross the line) and the variance of the residual series are lower than the univariate variances. Then, the multivariate predictions are closer than the univariate ones.

## IV.-RESULTS AND DISCUSSION

We have employed Equations 7 and 8 to model the cycle-to-cycle variability of $V_{set}$ and $V_{reset}$ as described in the previous section. Figure 5 shows the evolution of $V_{set}$ along a RS series we are considering here (1000 cycles), the corresponding univariate and multivariate models are also plotted for comparison. The multivariate model works better, as can be seen in the corresponding Figures (b)-(d) were different intervals of the RS series are selected.

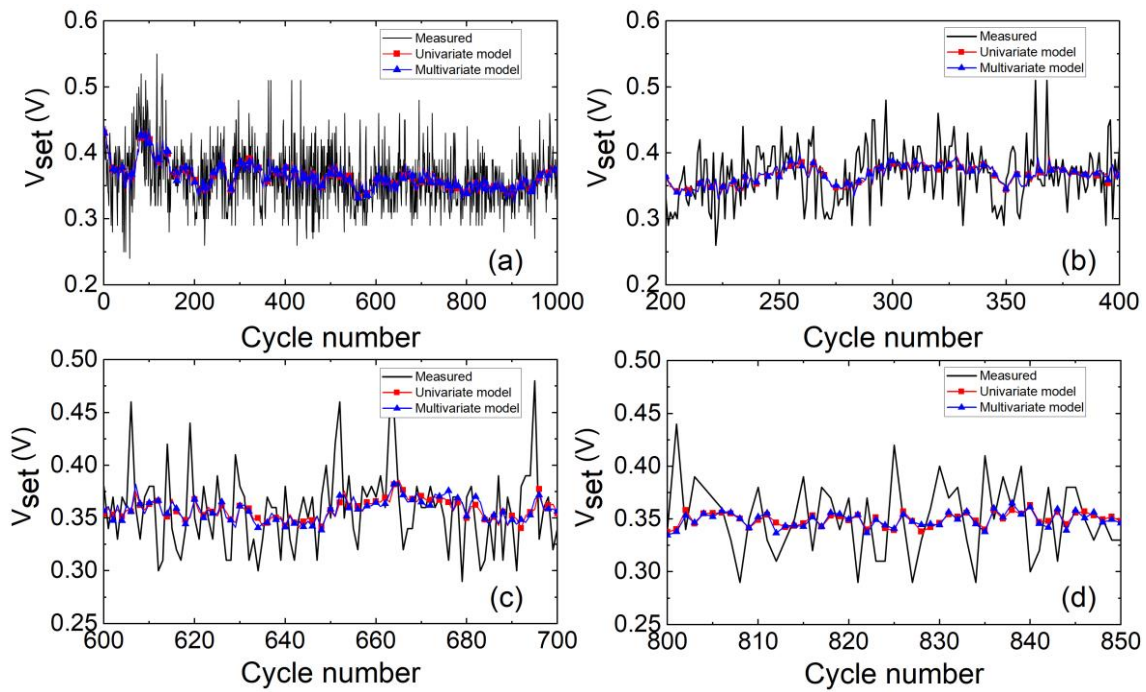

**Figure 5.** Experimental and modeled (in the univariate and multivariate cases) $V_{set}$ data for the whole RS series (a) and different cycle number intervals (b)-(d).

The accuracy of the multivariate approach was also reported in the previous section in terms of a comparison of the error variance. A similar conclusion can be noticed in Figure 6 where the $V_{reset}$ and the corresponding models are plotted.

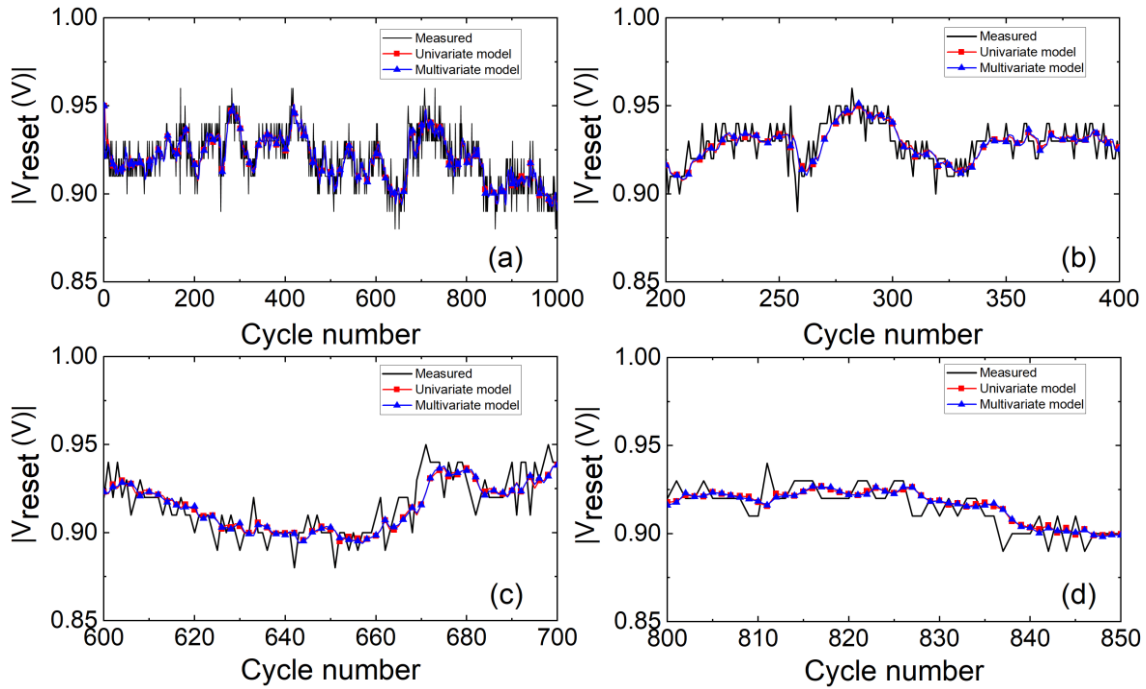

**Figure 6.** Experimental and modeled (in the univariate and multivariate cases) absolute value of $V_{reset}$ data for the whole RS series (a) and different cycle number intervals (b)-(d).

See in Figures 5 and 6 that the time series model designed here follows the RS series evolution of $V_{reset}$ and $V_{set}$ in a reasonable manner. The adequacy of this type of models to account for cycle-to-cycle variability for circuit simulation is remarkable since, to the best of our knowledge, no other modeling approaches can be obtained. There are other modeling options [26, 42], formulated in terms of means, variances and distribution functions, but no resistive switching series past values, or putting it into different words, resistive switching "inertia", are taken into consideration in the model, as it is the case here.

## V.-CONCLUSIONS

A new modeling methodology based on the time series analysis is introduced here to account for cycle-to-cycle variability in memristors. A multivariate approach is employed to model the reset and set voltages to enhance a univariate procedure previously developed. Experimental data from hafnium oxide based devices have been used for the model development; the modeled data are compared to experiments, a comparison of the univariate and multivariate approaches has been shown. The model accuracy is reasonable to be used in the context of circuit simulation to account for variability within resistive switching series linked to the inherent stochasticity of device operation.

## VI. - ACKNOWLEDGMENTS

We would like to thank F. Campabadal and M. B. González from the IMB-CNM (CSIC) in Barcelona for fabricating the devices employed here. The authors thank the support of the Spanish Ministry of Science, Innovation and Universities under projects TEC2017-84321-C4-3-R, MTM2017-88708-P, IJCI-2017-34038 (also supported by the FEDER program), the grant PGC2018-098860-B-I00 supported by MCIU/AEI/FEDER. This study has been partially financed by the Consejería de Conocimiento, Investigación y Universidad, Junta de Andalucía and European Regional Development Fund (ERDF) under projects A-TIC-117-UGR18 and A-FQM-345-UGR18. This work has made use of the Spanish ICTS Network MICRONANOFABS.